\documentclass[fleqn, preprint,12pt]{article}
\date{}

\usepackage[dvips]{graphicx}
\usepackage[cp 1251]{inputenc}
\usepackage{amsmath}
\usepackage{citesort}

\title{Pion photoproduction on nucleus with two-nucleon emission}
\author{Glavanakov  I.V., Tabachenko A.N.\\
\small \it Institute of Physics and Technology,  Tomsk Polytechnic  University, \\ \small \it 634050, Tomsk, Russia}

\begin{document}

\maketitle


\noindent
\parbox{15cm}

{A model for the pion photoproduction on nuclei in the $A\left( {\gamma ,\,
\pi NN} \right)B$ reaction is presented. This is an extension of our recent
model for the $A\left( {\gamma ,\, \pi N} \right)B$ reaction. In this
approach we have moved beyond the standard shell-model considering
$\Delta$\textit{N} correlations in the nuclear wave functions, which are caused by
the virtual transitions $NN \to \Delta N \to NN$ in the ground state of the
nucleus. The main ingredients of the model are the two- and three-particle
density matrices and the transition operators $\gamma \Delta \to N\pi $ and
$\gamma N \to N\pi $. The direct and exchange reaction mechanisms, which
follow from the structure of the density matrices, are examined. The model
is used to investigate the $^{12}$C$\left( {\gamma ,\, \pi ^{ +} p}
\right)$ and $^{12}$C$\left( {\gamma ,\, \pi ^{ -} p} \right)$
reactions in the kinematic region of the large momentum transfers to the
residual nuclear system, where the pion production occurs with the emission
of two nucleons. }



\section{Introduction}

In nuclear physics, one of the important problems is the study of the
nuclear structure at short and medium inter-nucleon distances, where the
effects of the non-nucleon degrees of freedom in nuclei appear. The short-
and middle-range structure of nuclei is defined by the short- and
middle-range components of the nucleon-nucleon potential, which lead to the
high-momentum components of the nuclear wave function. These wave function
components have a low probability. However, there are nuclear processes
which in the selected kinematic regions are almost completely caused by
these components of the wave function, which allows the various
manifestations of the non-nucleon degrees of freedom in nuclei to be
studied.

As is known, the two-nucleon knockout electromagnetic processes are a
powerful tool for studying the short- and middle- range dynamics of the
interaction between nucleons in nuclei.

In the framework of the independent particle model, the knocking-out from
the nucleus of two nucleons can occur by means of the two-body operators
corresponding to the meson exchange and isobar currents.

Another two-nucleon knockout mechanism is based on the nucleus model, in
which the correlated pairs of nucleons in the nucleus are taken into
account. This model is beyond the framework of the independent particle
model. The correlation of nucleons in the nucleus is described by the
correlation function, which reflects the structure of the nucleon-nucleon
potential. The process of knocking out nucleons in this model is due to the
action of the single-particle operator. As a result of knocking out either
nucleon, the second nucleon of the correlated pair can move to a free state.

Today these two approaches are widely used in the analysis of the knockout
$\left( {e,{e}'NN} \right)$ reactions, which is oriented mainly to the study
of short-range correlations induced by the repulsive part of the
nucleon-nucleon potential at small distances \cite{1}.

Another type of two-particle correlation in the nucleus is associated with
the virtual transitions $NN \to \Delta N \to NN$ in the ground state of the
nucleus. The interaction of the incident particle with the correlated
$\Delta N$ system may also lead to the knockout of two nucleons. These
$\Delta N$ correlations correspond to the middle-range components of the
nucleon-nucleon potential.

The role and relevance of these three competing processes can be different
in different reactions and kinematics. The peculiarity of manifestations of
the $\Delta$\textit{N}-correlations in nuclear reactions consists in the fact that
the knocking-out of the $\Delta$-isobar causes production of a pion as a result of
$\Delta \to N\pi $ decay. Therefore, because of the particle type in the
final state, the $\left( {\gamma ,\pi NN} \right)$ reactions are more
sensitive to the manifestations of the correlation of this type.

This article presents the analysis of the $A\left( {\gamma ,\pi NN}
\right)B$ process, taking into account the $\Delta$\textit{N}-correlations in the
ground state of the nucleus. The method of analysis is an extension of the
approach, developed in \cite{2,3} for the process $\left( {\gamma ,\pi N}
\right)$ at large momentum transfer, to pion photoproduction with the
two-nucleon knockout. The direct and exchange reaction mechanisms are
considered.

\section{Cross-section of the $A\left( {\gamma ,\pi NN} \right)B$ reaction}

The differential cross-section reaction
\[
A\left( {\gamma ,\pi NN} \right)B
\]
\noindent
can be written in the laboratory system of coordinates as
\[
d\sigma = 2\pi ^{}\delta \left( {E_{\gamma}  + M_{T} - E_{\pi}  - E_{1} -
E_{2} - E_{R}}  \right)^{}
\]
\[
\ \ \ \ \ \ \ \times \ \frac{{\overline {\left| {T_{fi}}  \right|^{2}}
}}{{4E_{\gamma}  E_{\pi} } }\frac{{d\textbf{p}_{\pi} } }{{\left( {2\pi}
\right)^{3}}}\frac{{d\textbf{p}_{1}} }{{\left( {2\pi}  \right)^{3}}}\frac{{d\textbf{p}_{2}
}}{{\left( {2\pi}  \right)^{3}}}\frac{{d\textbf{p}_{R}} }{{\left( {2\pi}
\right)^{3}}},
\]
\noindent
where (\textit{E}$_{\gamma}$, \textbf{p}$_{\gamma}$), (\textit{E}$_{\pi}$,
\textbf{p}$_{\pi}$), (\textit{E}$_{1}$, \textbf{p}$_{1}$), (\textit{E}$_{2}$,
\textbf{p}$_{2}$), and (\textit{E}$_{R}$, \textbf{p}$_{R}$) are the
four-momenta of the photon, pion, two nucleons, and the residual nucleus
\textit{B}; \textit{M}$_{T}$ is the mass of the nucleus \textit{A}; $T_{fi}
$ is the transition matrix element from the initial state, which includes
the photon and the nucleus \textit{A}, to the final, including the pion, two
nucleons in a free state and the residual nucleus \textit{B}.

The matrix element can be represented in the form
\[
T_{fi} = A\int {d\left( {X_{1} ',X_{1} ,X_{2} ,...,X_{A}}  \right)} _{} \Psi
_{F}^{\ast}  \left( {X_{1} ',X_{2} ,...,X_{A}}  \right)_{}
\]
\[
\ \ \ \ \ \ \ \times \  \left\langle
{X_{1} '} \right.\left| {t_{\gamma \pi} }  \right|\left. {X_{1}}
\right\rangle _{} \Psi _{T}^{} \left( {X_{1} ,X_{2} ,...,X_{A}}  \right).
\]
\noindent
Here $\Psi _{T} $ and $\Psi _{F} $ are the wave functions of the nucleus
\textit{A} and the system \textit{F}, which includes the free nucleons and
the residual nucleus \textit{B}; \textit{t}$_{\gamma\pi}$ is the single-particle
operator of the pion photoproduction on free baryons; \textit{X}$_{i}$ is
the coordinate in some space \textit{X}, fully characterizing the position
of the \textit{i}-th particle; and the integral means the summation over the
discrete and integration over the continuous variables.

Writing the wave function $\Psi _{F} $ as the antisymmetrized product of the
wave function $\phi _{\alpha _{1} \alpha _{2}}$, describing the state of
two free nucleons ($\alpha \equiv \textbf{p},m\sigma, m\tau$  is the nucleon state index),
and the wave function of the residual nucleus, we obtain
the following expression for $T_{fi} $
\[
T_{fi} = T_{d} - T_{e} ,
\]
\noindent
where
\[
 T_{d} = \left( {2A\left( {A - 1} \right)} \right)^{1/2}\int {d\left( {X_{1}
',X_{1} ,X_{2} ,...,X_{A}}  \right)_{}}  \phi _{\alpha _{1} \alpha _{2}
}^{\ast}  \left( {X_{1} ',X_{2}}  \right)^{}
\]
\[
 \ \ \ \ \ \ \ \ \times _{} \ \Psi _{f}^{\ast}  \left( {X_{3}
,X_{4} ,...,X_{A}}  \right)_{} \left\langle {X_{1} '} \right.\left| {t_{\gamma \pi} }
\right|\left. {X_{1}}  \right\rangle \Psi _{T}^{} \left( {X_{1} ,X_{2}
,...,X_{A}}  \right)
\]
\noindent
is the direct amplitude, in which the particle with the coordinate $X_{1} '$
is a free nucleon, and
\[
 T_{e} = \left( {\frac{{A\left( {A - 1} \right)}}{{2}}} \right)^{1/2}\left(
{A - 2} \right)\int {d\left( {X_{1} ',X_{1} ,X_{2} ,...,X_{A}}  \right)_{}}
\phi _{\alpha _{1} \alpha _{2}} ^{\ast}  \left( {X_{3} ,X_{2}}
\right)^{}
\]
\begin{equation}
\label{eq1}
 \ \ \ \ \ \ \ \ \times \  \Psi _{f}^{\ast}  \left( {X_{1} ',X_{4} ,...,X_{A}}  \right)
 \left\langle {X_{1} '} \right.\left| {t_{\gamma \pi} }
\right|\left. {X_{1}}  \right\rangle \Psi _{T}^{} \left( {X_{1} ,X_{2}
,...,X_{A}}  \right)
\end{equation}
\noindent
is the exchange amplitude, in which the particle with the coordinate $X_{1}
'$ is the part of the residual nucleus.

Consider the square of the modulus of the matrix element $T_{fi}$
\begin{equation}
\label{eq2}
\left| {T_{fi}}  \right|^{2} = \left| {T_{d}}  \right|^{2} + \left| {T_{e}}
\right|^{2} - 2Re\left( {T_{d} T_{e}^{\ast} }  \right).
\end{equation}

We are interested in differential cross-sections of the $A\left( {\gamma
,\pi NN} \right)B$ reaction, summed over the states of the residual nucleus.
Assuming that the set of states of the residual nucleus possesses the
completeness, the square of the modulus of the direct amplitude \textit{T}$_{d}$
can be expressed as
\[
 \sum\limits_{f} {\left| {T_{d}}  \right|^{2}} = 2A\left( {A - 1}
\right)\int {d\left( {X_{1} ',X_{1} ,X_{2} ,\tilde {X}_{1} ',\tilde {X}_{1}
,\tilde {X}_{2}}  \right)_{}} \phi _{\alpha _{1} \alpha _{2}} ^{\ast}
\left( {X_{1} ',X_{2}} \right)_{}
\]
\vspace*{-10pt}
\[
 \ \ \ \ \ \ \ \times _{}\   < X_{1} '|t_{\gamma \pi}  |X_{1} > \rho \left( {X_{1} ,X_{2}
;\tilde {X}_{1} ,\tilde {X}_{2}}  \right)
\]
\begin{equation}
\label{eq3}
 \ \ \ \ \ \ \  \ \ \ \ \ \ \  \times _{}\ < \tilde {X}_{1} |t_{\gamma \pi
}^{ +}  |\tilde {X}_{1} ' > \phi _{\alpha _{1} \alpha _{2}} ^{} \left(
{\tilde {X}_{1} ',\tilde {X}_{2}}  \right),
\end{equation}
\noindent
where
\[
\rho \left( {X_{1} ,X_{2} ;\tilde {X}_{1} ,\tilde {X}_{2}}  \right)
= \int {d\left( {X_{3} ,X_{4} ,...,X_{A}}  \right)_{}}
\]
\begin{equation}
\label{eq4}
\ \ \ \ \ \ \ \  \times _{}\  \Psi _{T}^{} \left( {X_{1}
,X_{2} ,X_{3} ,...,X_{A}}  \right)_{} \Psi _{T}^{\ast}  \left( {\tilde
{X}_{1} ,\tilde {X}_{2} ,X_{3} ,...,X_{A}}  \right)
\end{equation}
\noindent
is the two-particle density matrix.

Under the production of a charged pion by means of the exchange reaction
mechanism the "active" nucleon will most likely move to the level above the
Fermi level. In this case, the wave function of the residual nucleus can be
written as
\[
\Psi _{f}^{} \left( {X_{1} ',X_{4} ,...,X_{A}}  \right) = A_{u;1...A \ne
klm}^{s} \Psi _{\beta _{u}}  \left( {X_{1} '} \right)^{}\Psi _{\left( {\beta
_{k} \beta _{l} \beta _{m}}  \right)^{ - 1}} \left( {X_{4} ,...,X_{A}}
\right),
\]
\noindent
where the antisymmetrization operator $A_{u;1...A}^{s} $ rearranges the
indices of the particle states, $\beta _{u} $ is the index of the state of
the nucleon above the Fermi level, and $\left( {\beta _{k} \beta _{l} \beta
_{m}}  \right)^{ - 1}$ is the hole state of the bound system of particles
with the numbers 4, ..., \textit{A}. As a result, assuming that the set of
the hole states is complete, we obtain
\[
 \sum\limits_{f} {\left| {T_{e}}  \right|^{2}} = \frac{{A\left( {A - 1}
\right)\left( {A - 2} \right)}}{{2}}_{} \sum\limits_{u} {\int {d\left(
{X_{1} ',X_{1} ,X_{2} ,X_{3} ,\tilde {X}_{1} ',\tilde {X}_{1} ,\tilde
{X}_{2} ,\tilde {X}_{3}}  \right)_{}} }
\]
\[
 \ \ \  \times _{}\  \phi _{\alpha _{1} \alpha _{2}} ^{\ast}  \left( {X_{3} ,X_{2}}
\right)_{} \Psi _{\beta _{u}} ^{\ast}  \left( {X_{1} '} \right) < X_{1}
'|t_{\gamma \pi}  |X_{1} > \rho \left( {X_{1} ,X_{2} ,X_{3} ;\tilde {X}_{1}
,\tilde {X}_{2} ,\tilde {X}_{3}}  \right)_{}
\]
\begin{equation}
\label{eq5}
 \ \ \ \ \ \  \times _{} < \tilde {X}_{1} |
 t_{\gamma \pi} ^{ +}  |X_{1} ' > \Psi _{\beta
_{u}}  \left( {\tilde {X}_{1} '} \right)^{}\phi _{\alpha _{1} \alpha _{2}
}^{} \left( {\tilde {X}_{3} ,\tilde {X}_{2}}  \right).
\end{equation}
Here
\[
\rho \left( {X_{1} ,X_{2} ,X_{3} ;\tilde {X}_{1} ,\tilde {X}_{2} ,\tilde
{X}_{3}}  \right) = \int {d\left( {X_{4} ,...,X_{A}}  \right)} ^{}
\]
\[
\ \ \ \ \ \ \ \ \ \times _{}\ \Psi _{T}^{}
\left( {X_{1} ,X_{2} ,X_{3} ,...,X_{A}}  \right)^{}\Psi _{T}^{\ast}
\left( {\tilde {X}_{1} ,\tilde {X}_{2} ,\tilde {X}_{3} ,X_{4} ,...,X_{A}}
\right)
\]
\noindent
is the three-body density matrix.

Since the kinematic region where the main contributions of the direct and
exchange amplitudes are significantly different, in calculating the square
of the modulus $T_{fi} $, we neglect in (\ref{eq2}) the product
$T_{d} T_{e}^{\ast}$.

\section{The basic assumptions of the model}

We analyse the reaction $A\left( {\gamma ,_{} \pi NN} \right)B$ in the
framework of the formalism developed in \cite{4} for the description of the
ground state of nuclei, which previously we used in \cite{2,3} for considering
the reaction $A\left( {\gamma ,_{} \pi N} \right)B$ -- pion photoproduction
with the emission of a single nucleon. According to \cite{4}, baryons bound in
the nucleus, in addition to the space $\textbf{r}$, spin \textit{s}, and isospin
\textit{t} coordinates $\left( {r,s,t \equiv x} \right)$, are also
characterized by the intrinsic coordinate $m \quad \left( {x,m \equiv X}
\right)$. An eigenfunction $\Psi _{\beta}  \left( {X_{1} ,...,X_{A}}
\right)$ of the Hamiltonian \textit{H} of the system \textit{A} particles
with eigenvalue $E_{\beta}  $ is a superposition of the wave functions
concerned with different intrinsic configurations
\[
\Psi _{\beta}  \left( {X_{1} ,...,X_{A}}  \right) = \sum\limits_{n} {A_{n}}
\varphi _{n} \left( {m_{1} ,...,m_{A}}  \right)\Psi _{\beta} ^{n} \left(
{x_{1} ,...,x_{A}}  \right).
\]
Here $\Psi _{\beta} ^{n} \left( {x_{1} ,...,x_{A}}  \right)$ is the wave
function describing the state of \textit{A} particles in the usual, spin,
and isospin spaces; $\varphi _{n} \left( {m_{1} ,...,m_{{\rm A}}}  \right)$ is
the wave function describing the intrinsic states of baryons. The index
$\beta \equiv \beta _{1} ,...,\beta _{A} $ characterizes the usual space and
the spin and isospin states of \textit{A} particles. The index $n \equiv
n_{1} ,...,n_{A} $ defines the intrinsic states of the particles. For
instance, the state index describing the intrinsic configuration of the
nucleon system is written as $n = N_{1} ,N_{2} ,...,N_{A} $; if the first
particle is in the isobar state, but the rest are nucleons, the intrinsic
state index is written as $n = \Delta _{1} ,N_{2} ,...,N_{A} $. The wave
function $\Psi _{\beta} ^{n} \left( {x_{1} ,...,x_{A}}  \right)$ should be
antisymmetric for particles in the same intrinsic state. The remaining
antisymmetrization for particles in different intrinsic states is done by
the operator $A_{n} $. The wave functions $\varphi _{n} \left( {m_{1}
,...,m_{{\rm A}}}  \right)$ satisfy the condition
\[
\sum\limits_{m_{1} ,...,m_{A}}  {\varphi _{n} \left( {m_{1} ,...,m_{{\rm A}}
} \right)_{}}  \varphi _{\tilde {n}}^{\ast}  \left( {m_{1} ,...,m_{{\rm A}}
} \right) = \delta _{n_{1} ,\tilde {n}_{1}}  \cdot ... \cdot \delta _{n_{A}
,\tilde {n}_{A}}  .
\]

In our model, we will consider the two intrinsic configurations: a
configuration in which all the particles are nucleons and an isobar
configuration, in which one particle is $\Delta$-isobar and the others are
nucleons,
\[
 \Psi _{T} = \Psi _{T}^{N} + \Psi _{T}^{\Delta}  .
\]
Here $\Psi _{T}^{N} $ and $\Psi _{T}^{\Delta}  $ are the wave functions of
the nucleon and isobar configurations.

Assuming that only two nucleons are involved in the excitation of the
nucleon's internal degrees of freedom, the wave function $\Psi _{T}^{\Delta
} \left( {X_{1} ,...,X_{A}}  \right)$ of the isobar configuration can be
written as the superposition of the products of the wave function $\Psi
_{\left[ {\beta _{i} \beta _{j}}  \right]}^{\Delta N} \left( {X_{1} ,X_{2}}
\right)$ of the $\Delta$\textit{N} system, which includes an isobar and the second
nucleon (the participant of the transition \textit{NN} $\rightarrow \Delta$\textit{N}) and
the wave function $\Psi _{\left( {\beta _{i} \beta _{j}}  \right)^{ -
1}}^{N} \left( {X_{3} ,...,X_{A}}  \right)$, describing the state of the
nucleon core, which includes other \textit{A}--2 nucleons,
\begin{equation}
\label{eq6}
\Psi _{T}^{\Delta}  \left( {X_{1} ,...,X_{A}}  \right)^{} = ^{}A_{12;3...A}
\sum\limits_{ij} {\Psi _{\left[ {\beta _{i} \beta _{j}}  \right]}^{\Delta N}
\left( {X_{1} ,X_{2}}  \right)^{}\Psi _{\left( {\beta _{i} \beta _{j}}
\right)^{ - 1}}^{N} \left( {X_{3} ,...,X_{A}}  \right)} .
\end{equation}
Here
\[
A_{12;3...A} = \left( {\frac{{2}}{{A\left( {A - 1} \right)}}}
\right)^{1/2}\left[ {1 - \sum\limits_{i = 3}^{A} {\left( {P_{1i} + P_{2i}}
\right) + \sum\limits_{i = 3}^{A - 1} {\sum\limits_{j = i + 1}^{A} {P_{1i}
^{}P_{2j}} } } }  \right]
\]
\noindent
is the antisymmetrizaton operator; the operator $P_{ik} $ interchanges the
\textit{i}-th and \textit{k}-th nucleons,
\[
\Psi _{\left[ {\beta _{i} \beta _{j}}  \right]}^{\Delta N} \left( {X_{1}
,X_{2}}  \right)_{} = _{} A_{1;2} \ \varphi _{\Delta N} \left( {m_{1} ,m_{2}
} \right)_{} \Psi _{\left[ {\beta _{i} \beta _{j}}  \right]}^{\Delta N}
\left( {x_{1} ,x_{2}}  \right),
\]
\[
\Psi _{\left( {\beta _{i} \beta _{j}}  \right)^{ - 1}}^{N} \left( {X_{3}
,...,X_{A}}  \right) = \varphi _{N...N} \left( {m_{3} ,...,m_{A}}  \right)_{}
\Psi _{\left( {\beta _{i} \beta _{j}}  \right)^{ - 1}}^{N} \left( {x_{3}
,...,x_{A}}  \right).
\]

The oscillator shell model is used to describe the state of the \textit{A}--2
nucleons. The wave function of the $\Delta$\textit{N}-system is the solution of the
Schr$\ddot{o}$dinger equation for the potential due to the exchange of $\pi$- and
$\rho$-mesons, which describes the transition process $NN \to \Delta N$ \cite{4,5}.

According to (\ref{eq3}) and (\ref{eq5}) the square of the modulus of both the direct and
exchange amplitudes are expressed in terms of the density matrix and the
matrix elements of the single-particle operator of pion production
\textit{t}$_{\gamma\pi}$.

Considering the elementary processes, we take into account the reaction
mechanisms, which correspond to the single-particle transitions $\gamma N
\to N\pi $ and $\gamma \Delta \to N\pi $. We shall use the non-relativistic
operator of Blomqvist-Laget \cite{6} as the single-particle transition operator
$\gamma N \to N\pi $, which acts on the usual space, spin, and isospin
variables and is defined as
\[
 < x'|t_{\gamma N\pi} ^{} |x > _{} = \sum\limits_{{m}',m} {\varphi
_{N}^{\ast}  \left( {{m}'} \right)} < X'|t_{\gamma \pi} ^{} |X > \varphi
_{N} \left( {m} \right).
\]

Using the \textit{S}-matrix approach to the description of the $\gamma +
\Delta \to N + \pi $ processes, the transition operator $\gamma \Delta \to
N\pi $
\[
 < x'|t_{\gamma \Delta \pi} ^{} |x > _{} = \sum\limits_{{m}',m} {\varphi
_{N}^{\ast}  \left( {{m}'} \right)} < X'|t_{\gamma \pi} ^{} |X > \varphi
_{\Delta}  \left( {m} \right)
\]
was found. Let us to write it as an expansion on the four spin and three isospin
independent structures with the expansion coefficients, which depend on the
coupling constants and magnetic moments \cite{2}. The explicit form can be
written as
\[
 < x'|t_{\gamma \Delta \pi} ^{} |x > _{} = \delta \left( {{r}' - r}
\right)_{} e^{i_{} \left( {p_{\gamma}  r - p_{\pi}  {r}'} \right)} <
{s}',{t}'|t_{\gamma \Delta \pi} ^{} |s,t > .
\]
Here
\[
t_{\gamma \Delta \pi} ^{} = \phi _{a}^{ +}  \ \sum\limits_{i = 1}^{4}
{\sum\limits_{j = 1}^{3} {f_{ij} \ } }  M_{i} \  I_{j} ,
\]
\noindent
where $\phi _{a}^{} $ is the covariant unit vector of the cyclical basis
describing the isotopic state of the pion; index \textit{a} takes on the
values +, 0, and --, which fit with the positive, neutral, and negative
pions; $M_{i} $ are the independent spin structures
$$
\begin{array}{ll}
 M_{1}=\boldsymbol{\varepsilon}^{\lambda}\cdot\mathbf{S}^{+}; &
 M_{2}=i\boldsymbol{\sigma}\cdot[\mathbf{p}_{\gamma}\times\boldsymbol{\varepsilon}^{\lambda}]\ \mathbf{p}_{\pi}\cdot
\mathbf{S}^{+};
\\ & \\
 M_{3}=\mathbf{p}_{\pi}\cdot \boldsymbol{\varepsilon}^{\lambda}\ \mathbf{p}_{\gamma}\cdot
 \mathbf{S}^{+}; \ \ \ \ \ \ \ &
M_{4}=\mathbf{p}_{\pi}\cdot \boldsymbol{\varepsilon}^{\lambda}\ \mathbf{p}_{\pi}\cdot
 \mathbf{S}^{+}.
\end{array}
$$
Here $\boldsymbol{\varepsilon}^{\lambda}$ is the three-vector of the photon
polarization; $\boldsymbol{\sigma}$ is the Pauli matrix; $\mathbf{S}^{+}$ is the transition
spin operator, which converts the spin-3/2 state into the spin-1/2 state.

The isospin structures are
\[
\mathbf{I}_{1}=\mathbf{T}^{+},\ \mathbf{I}_{2}=\boldsymbol{\tau}T^{+}_{3},\
 \mathbf{I}_{3}=\tau_{3}\mathbf{T}^{+}
\],
\noindent
where $\boldsymbol{\tau}$ and $\mathbf{T}^{+}$ are analogues
$\boldsymbol{\sigma}$ and $\mathbf{S}^{+}$ in
isotopic space. Value $f_{ij} $ is a function of the photon and pion
momenta, the coupling constants $\pi NN$, $\pi N\Delta$, $\pi \Delta \Delta$,
the magnetic moments of the nucleon and $\Delta$-isobar, and the transition
magnetic moment $\mu _{\gamma N\Delta}  $. The explicit form of $f_{ij} $
and the used values of the coupling constants and the magnetic moments are
given in \cite{2}.

\section{Density matrices}

In this approach, according to (\ref{eq3}) and (\ref{eq5}), all information about the
structure of the nucleus and the mechanism of the $A\left( {\gamma ,\pi NN}
\right)B$ reaction is contained in the two- and three-body density matrices.

The two-particle density matrix $\rho \left( {X_{1} ,X_{2} ;\tilde {X}_{1}
,\tilde {X}_{2}}  \right)$ (\ref{eq4}) contained in the expression for the square
of the modulus of the direct amplitude (\ref{eq3}) of the \linebreak
$A\left( {\gamma ,\pi NN}
\right)B$ reaction was also used to describe the exchange mechanisms of the
pion production in the reaction $A\left( {\gamma ,\pi N} \right)B$ \cite{2}. We
are interested in the isobar configurations in the ground state of the
nucleus, so we consider the density matrix
\[
\rho ^{\Delta} \left( {X_{1} ,X_{2} ;\tilde {X}_{1} ,\tilde {X}_{2}}
\right)  = \int {d\left( {X_{3} ,X_{4} ,...,X_{A}}  \right)_{}}
\]
\[
\ \ \ \ \ \ \ \ \times \  \Psi
_{T}^{\Delta}  \left( {X_{1} ,X_{2} ,X_{3} ,...,X_{A}}  \right)_{} \Psi
_{T}^{\Delta \ast}  \left( {\tilde {X}_{1} ,\tilde {X}_{2} ,X_{3} ,...,X_{A}
} \right).
\]

Substituting in this expression the wave functions of the isobar
configuration presented in the form (\ref{eq6}), and integrating over $X_{3}
,...,X_{A} $, we obtain, according to \cite{2}
\[
\rho ^{\Delta} \left( {X_{1} ,X_{2} ;\tilde {X}_{1} ,\tilde {X}_{2}}
\right) = \rho _{\Delta N}^{} + \rho _{N\Delta} ^{} + \rho _{\Delta C}^{} +
\rho _{NC}^{} + \rho _{CN}^{} + \rho _{C\Delta} ^{} + \rho _{CC}^{} .
\]
Here, the two lower indices of the density matrix determine the state of the
particles with the numbers 1 and 2. Indices $\Delta$, \textit{N}, or \textit{C}
respectively indicate that the particle is a $\Delta$-isobar or a nucleon of the
$\Delta$\textit{N} system, or belongs to the nucleon core.

Because of the orthogonality of the wave functions $\varphi _{N} $ and
$\varphi _{\Delta}  $, the direct amplitudes corresponding to the matrices
$\rho _{N\Delta} ^{} $ and $\rho _{C\Delta} ^{} $ according to (\ref{eq3}) are zero.
In the amplitude corresponding to the $\rho _{CC}^{} $ matrix, the
$\Delta$\textit{N} system is a part of the residual nucleus and does not
dynamically manifest itself. This amplitude contributes to the cross-section
in a range of the low momentum transfer, where the quasi-free pion
photoproduction is dominant, so it will not be considered. The remaining
four terms $\rho _{\Delta N}^{}$, $\rho _{\Delta C}^{}$, $\rho _{NC}^{}
$, and $\rho _{CN}^{} $ of the two-particle density matrix
\[
\rho _{\Delta N}^{} \left( {X_{1} ,X_{2} ;\tilde {X}_{1} ,\tilde {X}_{2}}
\right)  = \varphi _{\Delta N} \left( {m_{1} ,m_{2}}  \right)
\]
\[
\ \ \ \ \ \ \times \ \left[
{\frac{{1}}{{A\left( {A - 1} \right)}}\sum\limits_{ij} {\Psi _{\left[ {\beta
_{i} \beta _{j}}  \right]}^{\Delta N} \left( {x_{1} ,x_{2}}  \right)^{}}
\Psi _{\left[ {\beta _{i} \beta _{j}}  \right]}^{\Delta N\ast}  \left(
{\tilde {x}_{1} ,\tilde {x}_{2}}  \right)} \right]\varphi _{\Delta N}^{\ast
} \left( {\tilde {m}_{1} ,\tilde {m}_{2}}  \right),
\]
\[
\rho _{\Delta C}^{} \left( {X_{1} ,X_{2} ;\tilde {X}_{1} ,\tilde {X}_{2}}
\right)
\]
\[
\ \ \ \ \ \  = \varphi _{\Delta N} \left( {m_{1} ,m_{2}}  \right)
\left[{\frac{{1}}{{A\left( {A - 1} \right)}}}\right.
\sum\limits_{ij,k \ne ij} {\int {d\left( {x_{3}}  \right)_{}}}
\Psi _{\left[ {\beta _{i} \beta _{j}}  \right]}^{\Delta N}
\left( {x_{1} ,x_{3}}  \right)_{}  \Psi _{\beta _{k}}
\left( {x_{2}}  \right)_{}
\]
\[
\ \ \ \ \ \ \ \ \ \ \ \ \left. {\times _{}\ \Psi _{\beta _{k}} ^{\ast}
\left( {\tilde {x}_{2}} \right)_{} \Psi _{\left[ {\beta _{i} \beta _{j}}\right]}
^{\Delta N\ast}  \left( {\tilde {x}_{1} ,x_{3}}  \right)} \right]
\varphi _{\Delta N}^{\ast} \left( {\tilde {m}_{1} ,\tilde {m}_{2}}  \right),
\]
\[
\rho _{NC}^{} \left( {X_{1} ,X_{2} ;\tilde {X}_{1} ,\tilde {X}_{2}}
\right)
\]
\[
\ \ \ \ \ \ =  \varphi _{NN} \left( {m_{1} ,m_{2}}  \right)
\left[ {\frac{{1}}{{A\left( {A - 1} \right)}}     }\right.
\sum\limits_{ij,k \ne ij} {\int {d\left( {x_{3}}  \right)^{}}
\Psi _{\left[ {\beta _{i} \beta _{j}}  \right]}^{\Delta N}
\left( {x_{3} ,x_{1}}  \right)^{}} \Psi _{\beta _{k}}
\left( {x_{2}}  \right)^{}
\]
\[
\ \ \ \ \ \ \ \ \ \ \ \ \left. {\times _{}\  \Psi _{\beta _{k}} ^{\ast}
\left( {\tilde {x}_{2}}  \right)^{}  \Psi _{\left[ {\beta _{i} \beta _{j}}
\right]}^{\Delta N\ast}  \left( {x_{3} ,\tilde {x}_{1}}  \right)} \right]
\varphi _{NN}^{\ast}  \left( {\tilde {m}_{1},\tilde {m}_{2}}  \right),
\]
\[
 \rho _{CN}^{} \left( {X_{1} ,X_{2} ;\tilde {X}_{1} ,\tilde {X}_{2}}
\right)
\]
\[
\ \ \ \ \ \  = \varphi _{NN} \left( {m_{1} ,m_{2}}  \right)
\left[ {\frac{{1}}{{A\left( {A - 1} \right)}}      }\right.
\sum\limits_{ij,k \ne ij} {\int {d\left( {x_{3}}  \right)^{}}
\Psi _{\left[ {\beta _{i} \beta _{j}}  \right]}^{\Delta N}
\left( {x_{3},x_{2}}  \right)^{}} \Psi _{\beta _{k}}
\left( {x_{1}}  \right)^{}
\]
\[
\ \ \ \ \ \ \ \ \ \ \ \ \left. {\times _{}\ \Psi _{\beta _{k}} ^{\ast}
\left( {\tilde {x}_{1}}  \right)^{}\Psi _{\left[ {\beta _{i} \beta _{j}}
\right]}^{\Delta N\ast}  \left( {x_{3} ,\tilde {x}_{2}}  \right)} \right]
\varphi _{NN}^{\ast}  \left( {\tilde {m}_{1},\tilde {m}_{2}}  \right)
\]
\noindent correspond to the reaction mechanisms, which are illustrated by the diagrams
shown in Fig. 1.
\begin{figure}[t]
\centering
\includegraphics [width = 10cm , keepaspectratio] {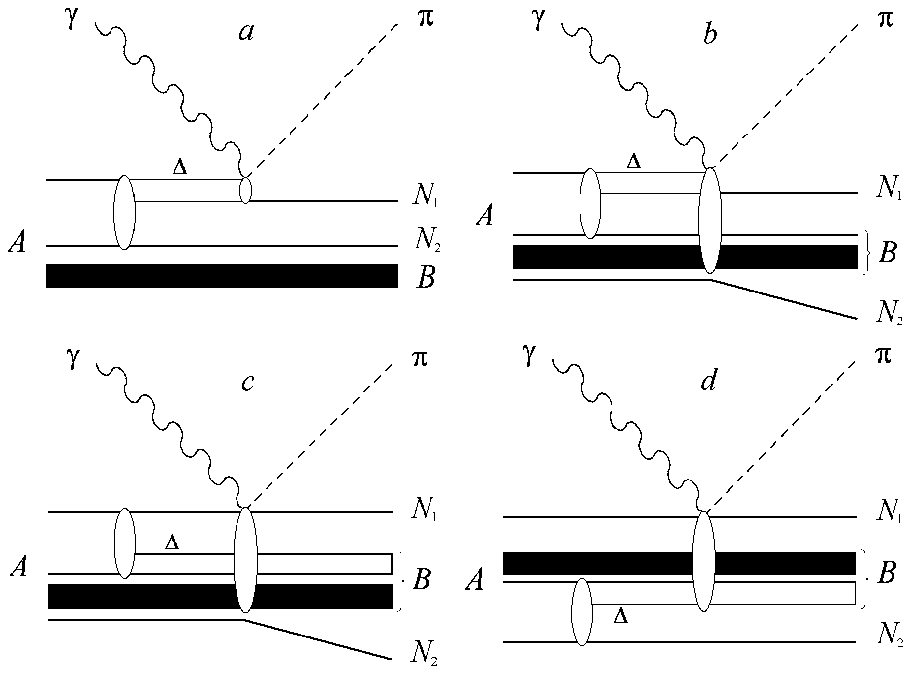}
\caption{Diagrams illustrating the direct mechanisms of the pion
photoproduction on nuclei in the $A\left( {\gamma ,\, \pi NN} \right)B$
reaction.}
\label{fig1}
\end{figure}

The correspondence between the individual terms of the density matrix and
diagrams in Fig. 1 is a simple one: the photon interacts with a baryon 1,
which together with a baryon 2 goes to a free state; baryons, over whose
coordinates the integration is carried out, are part of the residual
nucleus. Each term of the density matrix corresponds to the appointed final
state, which depends on the structure of the matrix and the $ < X_{1}
'|t_{\gamma \pi}  |X_{1} > $ operator.

We now consider the three-particle density matrix
\[
\rho ^{\Delta} \left( {X_{1} ,X_{2} ,X_{3} ;\tilde {X}_{1} ,\tilde {X}_{2}
,\tilde {X}_{3}}  \right) = \int {d\left( {X_{4} ,...,X_{A}}  \right)_{}}
\]
\[
\ \ \ \ \ \ \ \ \  \times \
\Psi _{T}^{\Delta}  \left( {X_{1} ,X_{2} ,X_{3} ,X_{4} ,...,X_{A}}  \right)_{} \Psi
_{T}^{\Delta \ast}  \left( {\tilde {X}_{1} ,\tilde {X}_{2} ,\tilde {X}_{3}
,X_{4} ,...,X_{A}}  \right).
\]
The expression for $\rho ^{\Delta}$ can be represented as a
sum of five terms, as a result of the transformation, similar to that
carried out with the two-particle density matrix
\[
\rho ^{\Delta}  = \rho _{CSS}^{} + \rho _{SSC}^{} + \rho _{SCC}^{}
+ \rho _{CSC}^{} + \rho _{CCC}^{} ,
\]
 where
\[
\rho _{CSS}^{} \left( {X_{1} ,X_{2} ,X_{3} ;\tilde {X}_{1} ,\tilde {X}_{2}
,\tilde {X}_{3}}  \right) = \frac{{2}}{{A\left( {A - 1} \right)\left( {A -
2} \right)}}
\]
\[
\ \ \ \ \ \ \ \ \times \  \sum\limits _{ij,k \ne ij} {\Psi _{\beta _{k}}  \left( {X_{1}}  \right)_{}
\Psi _{\left[ {\beta _{i} \beta _{j}}  \right]}^{\Delta N} \left( {X_{2}
,X_{3}}  \right)_{}}  \ \Psi _{\left[ {\beta _{i} \beta _{j}}
\right]}^{\Delta N\ast}  \left( {\tilde {X}_{2} ,\tilde {X}_{3}}  \right)_{}
\Psi _{\beta _{k}} ^{\ast}  \left( {\tilde {X}_{1}}  \right),
\]
\[
\rho _{SSC}^{} \left( {X_{1} ,X_{2} ,X_{3} ;\tilde {X}_{1} ,\tilde {X}_{2}
,\tilde {X}_{3}}  \right) = \frac{{4}}{{A\left( {A - 1} \right)\left( {A -
2} \right)}}{}_{}
\]
\[
\ \ \ \ \ \ \ \  \times \ \sum\limits_{ij,k \ne ij} {\Psi _{\beta _{k}}
\left( {X_{3}}  \right)_{}
\Psi _{\left[ {\beta _{i} \beta _{j}}  \right]}^{\Delta N} \left( {X_{1}
,X_{2}}  \right)_{}}  \Psi _{\left[ {\beta _{i} \beta _{j}}
\right]}^{\Delta N\ast}  \left( {\tilde {X}_{1} ,\tilde {X}_{2}}  \right)_{}
\Psi _{\beta _{k}} ^{\ast}  \left( {\tilde {X}_{3}}
\right),
\]
\[
\rho _{SCC}^{} \left( {X_{1} ,X_{2} ,X_{3} ;\tilde {X}_{1} ,\tilde {X}_{2}
,\tilde {X}_{3}}  \right) = \frac{{4}}{{A\left( {A - 1} \right)\left( {A -
2} \right)}}
 \sum\limits_{ij,kl \ne ij} {\Psi _{\beta _{k} \beta _{l}}  \left( {X_{2} ,X_{3}}  \right)^{}}
\]
\[
\ \ \ \ \ \ \ \  \times \  {\int {d\left( {X_{4}}  \right)} _{} \Psi _{\left[ {\beta
_{i} \beta _{j}}  \right]}^{\Delta N} \left( {X_{1} ,X_{4}}  \right)^{}}
\Psi _{\left[ {\beta _{i} \beta _{j}}  \right]}^{\Delta N\ast}  \left(
{\tilde {X}_{1} X_{4}}  \right)^{}\Psi _{\beta _{k} \beta _{l}} ^{\ast}
\left( {\tilde {X}_{2} ,\tilde {X}_{3}}  \right),
\]
\[
 \rho _{CSC}^{} \left( {X_{1} ,X_{2} ,X_{3} ;\tilde {X}_{1} ,\tilde {X}_{2}
,\tilde {X}_{3}}  \right) = \frac{{8}}{{A\left( {A - 1} \right)\left( {A -
2} \right)}}{}_{} \sum\limits_{ij,kl \ne ij}
{\Psi _{\beta _{k} \beta _{l}}  \left( {X_{1}
,X_{3}}  \right)^{}}
\]
\[
\ \ \ \ \ \ \ \  \times \   {\int {d\left( {X_{4}}  \right)} _{} \Psi _{\left[ {\beta
_{i} \beta _{j}}  \right]}^{\Delta N} \left( {X_{2} ,X_{4}}  \right)^{}}
\Psi _{\left[ {\beta _{i} \beta _{j}}  \right]}^{\Delta N\ast}  \left(
{\tilde {X}_{2} ,X_{4}}  \right)^{}\Psi _{\beta _{k} \beta _{l}} ^{\ast}
\left( {\tilde {X}_{1} ,\tilde {X}_{3}}  \right),
\]
\[
\rho _{CCC}^{} \left( {X_{1} ,X_{2} ,X_{3} ;\tilde {X}_{1} ,\tilde {X}_{2}
,\tilde {X}_{3}}  \right) = \frac{{6}}{{A\left( {A - 1} \right)\left( {A -
2} \right)}}{}_{}
\]
\[
\ \ \ \ \ \ \ \  \times \ \sum\limits_{ij,klm \ne ij}
\Psi _{\beta _{k} \beta _{l} \beta _{m}} \left( {X_{1}
,X_{2} ,X_{3}}  \right)^{}\int {d\left( {X_{4} ,X_{5}}
\right)} _{} \Psi _{\left[ {\beta _{i} \beta _{j}}  \right]}^{\Delta N}
\]
\[
\ \ \ \ \ \ \ \ \ \ \ \ \ \ \ \ \Psi _{\left[ {\beta _{i} \beta _{j}}
\right]}^{\Delta N\ast}  \left( {X_{4} ,X_{5}}  \right)_{}
 \Psi _{\beta _{k} \beta _{l} \beta _{m}}
\left( {\tilde {X}_{1} ,\tilde {X}_{2} ,\tilde {X}_{3}}  \right).
\]

 Summing these formulas over the internal variables, we get the density
matrix in the form of a sum of nine terms
\[
\rho ^{\Delta} \left( {X_{1} ,X_{2} ,X_{3} ;\tilde {X}_{1} ,\tilde {X}_{2}
,\tilde {X}_{3}}  \right)
\]
\[
\ \ \ \ \ \ \ \ = \rho _{C\Delta N}^{} + \rho _{CN\Delta} ^{} +
\rho _{\Delta NC}^{} + \rho _{N\Delta C}^{} + \rho _{\Delta CC}^{} + \rho
_{NCC}^{} + \rho _{C\Delta C}^{} + \rho _{CNC}^{} + \rho _{CCC}^{} .
\]
Here the subscripts define the state of the particles with the numbers 1, 2,
and 3.

Recall that the exchange amplitude (\ref{eq1}) is presented in the form in which the
photon interacts with particle 1 while particles 2 and 3 go to a free state.
Because of the orthogonality of the internal wave functions, the exchange
amplitudes of the $A\left( {\gamma ,\,\pi NN} \right)B$ process,
corresponding to the matrices $\rho _{C\Delta N}^{} ,\ \rho _{CN\Delta} ^{}
,\ \rho _{N\Delta C}^{} ,\ \rho _{C\Delta C}^{} $, are zero. Amplitudes
associated with the matrices $\rho _{\Delta CC}^{} ,\ \rho _{NCC}^{} $,
and $\rho _{CCC}^{} $ answer to the reaction mechanism, in which
the nucleons in a state lower than the Fermi level go to a free state. The
probability of such processes with the nucleon momentum in excess of $\sim $
200 MeV/c is very small. Therefore, we ignore them. The remaining two
matrices
\
\[
 \rho _{\Delta NC}^{} \left( {X_{1} ,X_{2} ,X_{3} ;\tilde {X}_{1} ,\tilde
{X}_{2} ,\tilde {X}_{3}}  \right) = \frac{{2}}{{A\left( {A - 1}
\right)\left( {A - 2} \right)}}{}_{}
\]
\[
\ \ \ \ \ \ \ \ \times \ \varphi _{\Delta NN} \left( {m_{1} m_{2} m_{3}}  \right)^{}\left[
{\sum\limits_{ijk \ne ij}} {\Psi _{\beta _{i} \beta _{j}} ^{\Delta N} \left(
{x_{1} x_{2}}  \right)^{}\Psi _{\beta _{k}}  \left( {x_{3}}  \right)^{}}  \right.
\]
\[
\ \ \ \ \ \ \ \ \ \ \ \ \ \ \ \ \times \ \left. \Psi _{\beta _{k}} ^{\ast}
\left( {\tilde {x}_{3}}  \right)^{}\Psi _{\beta
_{i} \beta _{j}} ^{\Delta N\ast}  \left( {\tilde {x}_{1} \tilde {x}_{2}}
\right) \right]{}^{}\varphi _{\Delta NN}^{\ast}  \left( {\tilde {m}_{1}
\tilde {m}_{2} \tilde {m}_{3}}  \right),
\]
\[
 \rho _{CNC}^{} \left( {X_{1} ,X_{2} ,X_{3} ;\tilde {X}_{1} ,\tilde {X}_{2}
,\tilde {X}_{3}}  \right) = \frac{{4}}{{A\left( {A - 1} \right)\left( {A -
2} \right)}}{}_{}
\]
\[
\ \ \ \ \ \ \ \ \times \ \varphi _{NNN} \left( {m_{1} m_{2} m_{3}}  \right)\left[  \int {d\left(
{x_{4}}  \right)} \sum\limits _{ij,kl \ne ij} {\Psi _{\beta _{i} \beta _{j}
}^{\Delta N} \left( {x_{4} x_{2}}  \right)^{}\Psi _{\beta _{k} \beta _{l}}
\left( {x_{1} x_{3}}  \right)^{}} \right.
\]
\[
\ \ \ \ \ \ \ \ \ \ \ \ \ \ \ \ \times \ \left. \Psi _{\beta _{k} \beta _{l}} ^{\ast}
\left( {\tilde {x}_{1} \tilde {x}_{3}}  \right)_{}
\Psi _{\beta _{i} \beta _{j}} ^{\Delta N\ast}  \left( {x_{4} \tilde {x}_{2}}  \right) \right]
\varphi _{NNN}^{\ast}  \left( {\tilde {m}_{1} \tilde {m}_{2} \tilde {m}_{3}}  \right)
\]
correspond to the reaction mechanisms, which are illustrated by the diagrams
in Fig. 2. Common to these two matrices is that the nucleon of the
$\Delta$\textit{N} system goes to a free state.

\section{Analysis of experimental data}

At present, the experimental data of the $A\left( {\gamma ,\, \pi NN} \right)B$
reaction are practically absent. Therefore, to compare the predictions of
our model with experimental data, we will use the $A\left( {\gamma ,\, \pi N}
\right)$ reaction measured in the kinematic region where, according to \cite{7},
the pion production occurs in the $A\left( {\gamma ,\, \pi N} \right)NB$
reaction with the emission of two nucleons. This kinematic region is
characterized primarily by the large momentum transfers to the \textit{NB}
system, consisting of the free nucleon and the residual nucleus.

\begin{figure}[t]
\centering
\includegraphics [width = 10cm , keepaspectratio] {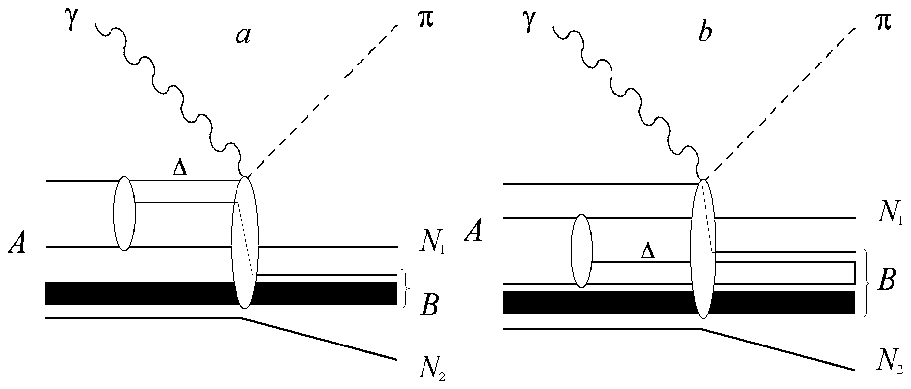}
\caption{Diagrams illustrating the exchange mechanisms of the pion
photoproduction on nuclei in the $A\left( {\gamma ,\, \pi NN} \right)B$
reaction.}
\label{fig2}
\end{figure}

We will consider the data from two experiments carried out at the Tomsk
synchrotron in which the ${}^{12}$C$\left( {\gamma ,\, \pi ^{ -} p}
\right)$ \cite{7} and ${}^{12}$C$\left( {\gamma ,\, \pi ^{ +} p} \right)$
\cite{8} reactions were examined. Both experiments were performed in the
kinematic region of the large momentum transfers to the residual system and
have repeatedly attracted attention \cite{2,3,9,10}. The data of the
${}^{12}$C$\left( {\gamma ,\, \pi ^{ -} p} \right)$ reactions are
interesting in that at large opening angles of the pion-proton pair the
maximum of the cross-section is observed, which has been interpreted as a
manifestation of a quasi-bound isobar-nuclear state -- a highly excited state
of the nucleus which decays with the emission of the pion and nucleon
\cite{7,9,10}. The second ${}^{12}$C$\left( {\gamma ,\, \pi ^{ +} p}
\right)$ reaction, which is forbidden for the quasi-free pion
photoproduction mechanism, is useful to study the isobar configuration in
the ground state of the nucleus.

The experimental point in Fig. 3 is the differential cross-section, measured
by using the bremsstrahlung beam of electrons in the two runs with electron
beam energies of 500 and 420 MeV \cite{8}. The measurements were performed by the
simultaneous detection of a positive pion and a proton. Positive pions were
detected at an angle of 54$^\circ$ to the photon beam axis. In Fig. 3 both the
experimental data and the calculated cross-section are averaged over the
proton energy in the interval of 80-120 MeV, over the pion energy in the
interval of 71.5-106.5 MeV, and over the proton polar angle in the interval
of 56-94$^\circ$. In the kinematic region under consideration, the average photon
energy and the average residual nucleus momentum were 355 MeV and 300
MeV/\textit{c}, respectively. Supposing that the pion production is in the
${}^{12}$C$\left( {\gamma ,\, \pi ^{ +} p} \right)^{11}$Be reaction,
the cross-section was calculated on the basis of data on the differential
yield of the reaction.

The theoretical differential cross-section of the ${}^{12}$C$\left( {\gamma
,\, \pi ^{ +} p} \right)^{11}$Be reaction depending on the energy of
the proton is shown in Fig. 3 by a dashed-dotted curve. Using a model that
takes into account the isobar configuration in the ground state of the
nucleus, the calculations are performed. The wave function of the nucleon
bound state is calculated using the harmonic-oscillator shell model which
reproduces the charge radius of the $^{12}$C nucleus. Final-state
interaction is taken into account through an optical model. The model is
described in detail in \cite{2,3}. As can be seen, the calculated cross-section
of the ${}^{12}$C$\left( {\gamma ,\, \pi ^{ +} p} \right)^{11}$Be
reaction is about $ \sim $0.1 of the experimental cross-section.
\begin{figure}[t]
\centering
\includegraphics [width = 8cm , keepaspectratio] {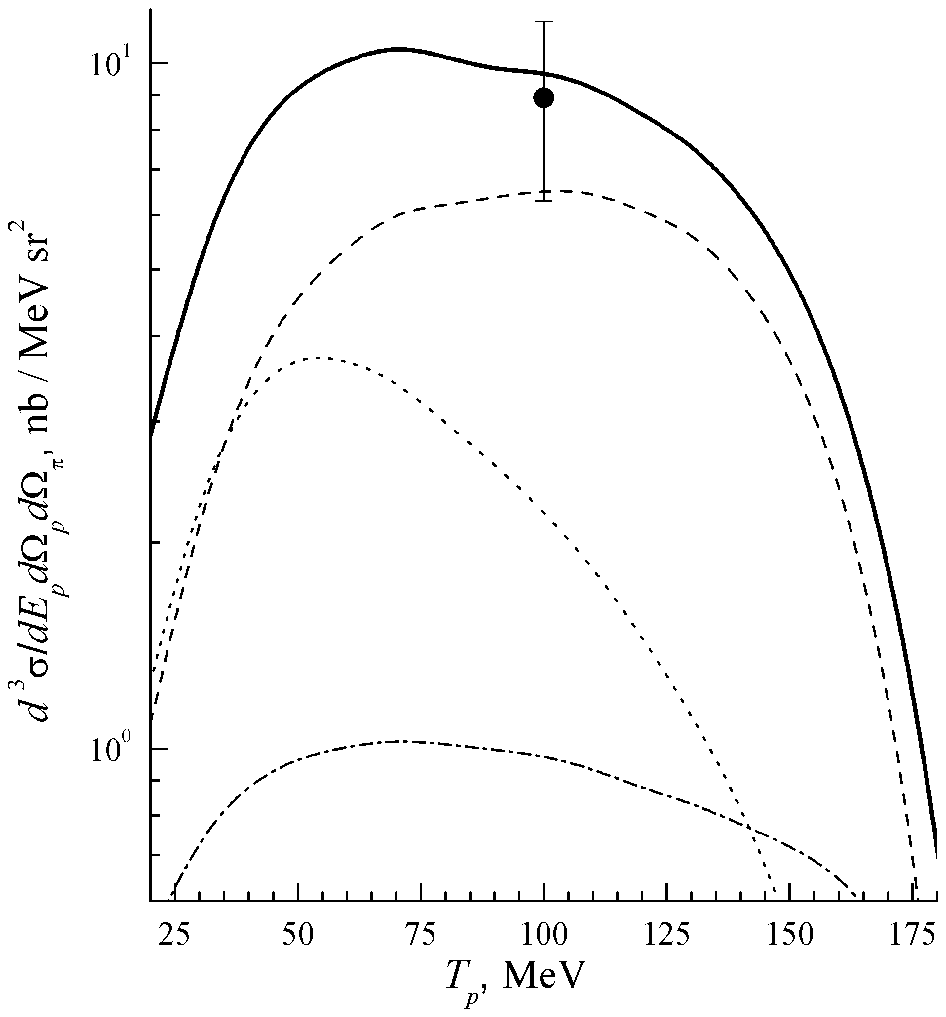}
\caption{Differential cross-section of the ${}^{12}$C$\left( {\gamma ,\,
\pi ^{ +} p} \right)$ reaction for $\bar {E}_{\gamma}  $ = 355 MeV versus
the kinetic energy of the proton \textit{T}$_{p}$. Dashed-dotted curve:
cross-section of the${}^{12}$C$\left( {\gamma ,\, \pi ^{ +} p}
\right)^{11}Be$ reaction; dashed curve: summed contribution to the
${}^{12}$C$\left( {\gamma ,\, \pi ^{ +} p} \right)NB$ reactions from
the mechanisms, corresponding to the diagrams in Figs. 1\textit{a} and
1\textit{b}; dotted curve: summed contribution to the ${}^{12}$C$\left(
{\gamma ,\, \pi ^{ +} p} \right)NB$ reactions from the mechanisms,
corresponding to the diagrams in Figs. 1\textit{d}, 2\textit{a}, and
2\textit{b}; solid curve: total contribution to the cross-section of the
${}^{12}$C$\left( {\gamma ,\, \pi ^{ +} p} \right)^{11}$Be,
${}^{12}$C$\left( {\gamma ,\, \pi ^{ +} p} \right)n^{10}$Be, and
${}^{12}$C$\left( {\gamma ,\, \pi ^{ +} p} \right)p^{10}$Li reactions;
data are taken from Ref. \cite{8}.}
\label{fig3}
\end{figure}

The dashed and dotted curves in Fig. 3 show the $d^{3}\tilde {\sigma} _{\pi
N} $ contributions to the experimental cross-section of the ${}^{12}$C$\left(
{\gamma ,\, \pi ^{ +} p} \right)$ reaction of the events from the
${}^{12}$C$\left( {\gamma ,\, \pi ^{ +} p} \right)NB$ reaction, in
which the residual nuclear system \textit{NB} contains a nucleon in a free
state. The contribution $d^{3}\tilde {\sigma} _{\pi N} $ is related with the
cross-section $d^{5}\sigma _{\pi NN} $ of the ${}^{12}$C$\left( {\gamma ,\,
 \pi ^{ +} p} \right)NB$ reaction by the relation
\[
\frac{{d^{3}\tilde {\sigma} _{\pi N}} }{{dE_{p} d\Omega _{p} d\Omega _{\pi}
}}\left| {\frac{{\partial E_{\gamma} } }{{\partial E_{\pi} } }}
\right|{}^{}f\left( {E_{\gamma} }  \right) = \int {dE_{\gamma}^{'} }  d\Omega
_{N} {}^{}f\left( {E_{\gamma}^{'} }  \right)\frac{{d^{5}\sigma _{\pi NN}
}}{{dE_{p} d\Omega _{p} dE_{\pi}  d\Omega _{\pi}  d\Omega _{N}} },
\]
where $f\left( {E_{\gamma} }  \right)$ is the bremsstrahlung spectrum of the
electrons. The kinematic variables in the left side of this formula satisfy
the law of energy and momentum conservation in the ${}^{12}$C$\left( {\gamma
,\, \pi ^{ +} p} \right)^{11}$Be reaction.

In the chosen kinematic region, the main factors determining the proton
energy dependence of the cross-section of the ${}^{12}$C$\left( {\gamma ,\,
 \pi ^{ +} p} \right)NB$ reaction are the momentum distributions of the
isobar and the nucleon of the $\Delta$\textit{N} system in the nucleus. The
influence that these factors have on the cross-section depends on the
reaction mechanism. Dashed and dotted curves in Fig. 3 show the summed
cross-section contributions from the reaction mechanisms, corresponding to
the two groups of diagrams. The first group includes the diagrams in Figs.
1\textit{a} and 1\textit{b}, and the second, the diagram in Figs.
1\textit{d}, 2\textit{a}, and 2\textit{b}.
The main difference between
these diagrams is that in the first group the proton of the $\pi N$-pair is the
product of the $\gamma \Delta ^{ + +}  \to \pi ^{ +} p$ transition while in
the second group the proton is the nucleon of the $\Delta$\textit{N} system. It
should be noted that the amplitude of the reaction, corresponding to the
diagram in Fig. 1\textit{a}, includes both the $\gamma \Delta ^{ + +}  \to
\pi ^{ +} p$ transition and the $\gamma \Delta ^{ +}  \to \pi ^{ +} n$
transition. However, the $\gamma \Delta ^{ + +}  \to \pi ^{ +} p$ transition
is dominant. The position of the cross-section maximum, represented by the
dashed and dotted curves, is determined mainly by the dependence on the
momentum of the Fourier transform of the wave function
$\Psi ^{\Delta N} $, describing the relative motion of
the nucleon and the isobar of the $\Delta$\textit{N} system, which has a maximum at
a momentum of 320 MeV/\textit{c}. The solid curve in Fig. 3 shows the total
contribution to the cross-section of the ${}^{12}$C$\left( {\gamma ,\,
\pi ^{ +} p} \right)^{11}$Be, ${}^{12}$C$\left( {\gamma ,\, \pi ^{ +
}p} \right)n^{10}$Be, and ${}^{12}$C$\left( {\gamma ,\, \pi ^{ +} p}
\right)p^{10}$Li reactions. In the kinematic region under consideration the
mechanism corresponding to the diagram in Fig. 1\textit{b} dominates. As can
be seen, the calculated cross-section is in good agreement with the
experimental data.

Figure 4 shows the differential yield of the ${}^{12}$C$\left( {\gamma ,\,
 \pi ^{ -} p} \right)$ reaction depending on the opening angle $\Theta
_{\pi p} $ of the pion and proton \cite{7}. Experimental data were obtained under
the following conditions. Pions with momentum 224 MeV/\textit{c} were detected at an
angle of 76$^\circ$ with respect to the axis of the photon beam. The measurement
results are averaged over the proton energy range 60-140 MeV. The experiment
was performed at the bremsstrahlung beam of the electrons with an energy of
500 MeV.

The short dashed curve in Fig. 4, representing the yield of the
reaction ${}^{12}$C$\left( {\gamma ,\, \pi ^{ -} p} \right)^{11}$C in
the quasi-free approximation \cite{11}, satisfactorily describes the exponential
decrease of the reaction yield with increases in the opening angle up to
$\Theta _{\pi p} \simeq 150^\circ $. With further increases in $\Theta _{\pi
p} $, a sharp disagreement takes place between the experimental and
calculated data.
\begin{figure}[t]
\centering
\includegraphics [width = 10cm , keepaspectratio] {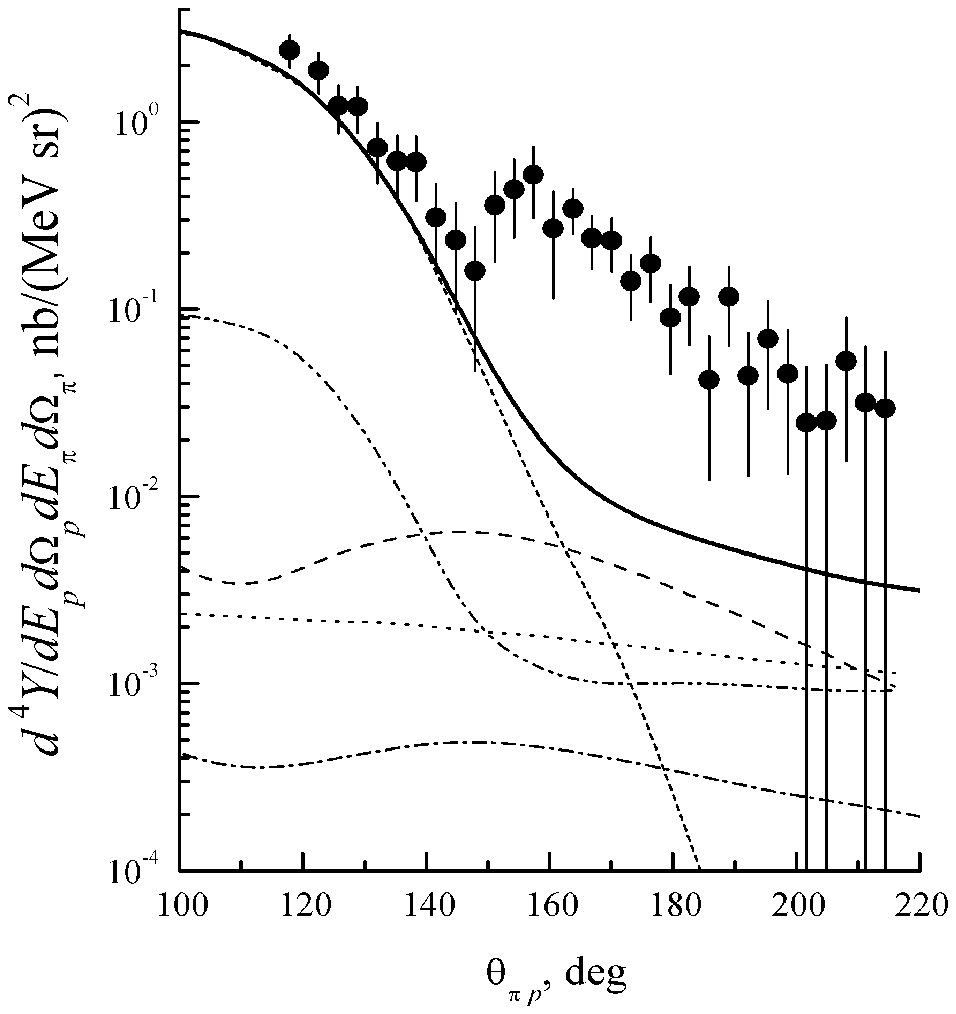}
\caption{Differential yield of the $^{12}$C$\left(\gamma ,\, \pi^{-} p \right)$
reaction versus the $\Theta _{\pi p}$ opening angle. Short
dashed curve: quasi-free pion photoproduction in the
$^{12}$C$\left(\gamma,\, \pi ^{-} p \right)^{11}$C reaction; dashed-dotted curve: the
contribution of the isobar configuration in the $^{12}$C ground state to the
yield of the $^{12}$C$\left(\gamma ,\, \pi ^{-} p \right)^{11}$C
reaction; dashed curve: summed contribution from the mechanisms of the
$^{12}$C$\left(\gamma ,\, \pi ^{-} p \right)NB$ reactions,
corresponding to the diagrams in Figs. 1\textit{b} and 1\textit{c}; dotted
curve: summed contribution from the mechanisms of the
$^{12}$C$\left(\gamma ,\, \pi ^{-} p \right)NB$ reactions, corresponding to the
diagrams in Figs. 1\textit{a}, 2\textit{a}, and 2\textit{b};
dashed-dotted-dotted curve: the contribution from the mechanism of the
$^{12}$C$\left(\gamma ,\, \pi ^{-} p \right)NB$ reactions,
corresponding to the diagram in Fig. 1\textit{d}; solid curve: total
contribution of the $^{12}$C$\left(\gamma ,\, \pi ^{-} p
\right)^{11}$C, $^{12}$C$\left(\gamma ,\, \pi ^{-} p
\right)n^{10}$C, and $^{12}$C$\left(\gamma ,\, \pi ^{-} p \right)p^{10}$B
reactions; data are taken from Ref. \cite{7}.}
\label{fig4}
\end{figure}
The dashed-dotted curve in Fig. 4 shows the contribution of the isobar
configuration in the ground state of $^{12}$C to the yield of the
${}^{12}$C$\left( {\gamma ,\, \pi ^{ -} p} \right)^{11}$C reaction,
which at a high opening angle is not more than $10^{ - 2}$ of the
experimental cross-section \cite{2}. The dashed, dotted, and dash-dot-dot curves
represented in Fig. 4 are the sums of the contributions from the mechanisms
of the ${}^{12}$C$\left( {\gamma ,\, \pi ^{ -} p} \right)NB$ reaction,
corresponding to the three groups of diagrams, which lead to the
cross-sections with the different angular correlations.

The first group includes the diagrams in Figs. 1\textit{b} and 1\textit{c}.
The contribution to the reaction yield of these diagrams is shown in Fig. 4
by the dashed curve. In this case, the angular correlation of the reaction
yield is due to the strong dependence of the momentum of the pion-proton
pairs and, therefore, the active baryon momentum in the initial state $\textbf{p}_{B}
= \textbf{p}_{\pi}  + \textbf{p}_{p} - \textbf{p}_{\gamma}  $ from the opening angle $\Theta _{\pi p}
$ of the pion and proton. The maximum of the angular dependence of the
reaction yield, associated with the first group of diagrams, has the same
nature as the maximum of the cross-section of the ${}^{12}$C$\left( {\gamma
,\, \pi ^{ +} p} \right)NB$ reaction, shown in Fig. 3 by the dashed
curve. At the opening angle $\Theta _{\pi p} \simeq 150^\circ $ the mean
momentum is $\left| {p_{B}}  \right| \simeq 320$ MeV/\textit{c}. The second
group includes the diagrams in Figs. 1\textit{a}, 2\textit{a}, and
2\textit{b}. The diagram in Fig. 1\textit{a} contains $\gamma \Delta ^{0}
\to \pi ^{ -} p$ and $\gamma \Delta ^{ -}  \to \pi ^{ -} n$ transitions. The
last transition, at which the proton is a nucleon of the $\Delta$\textit{N} system,
dominates. Therefore, we have included this diagram in the second group. In
the amplitudes corresponding to the diagrams of the second group, the escape
directions of the pion and proton are dynamically weakly correlated. The
reaction yield contribution of this group's diagrams is shown in Fig. 4 by
the dotted curve. The observed slight decrease in yield of the reaction with
increasing opening angle mainly has the phase-space nature.

The third group includes the diagram in Fig. 1\textit{d}. The contribution
of this diagram to the yield of the ${}^{12}$C$\left( {\gamma ,\, \pi
^{ -} p} \right)n{}^{10}$C$\quad $ reaction is concentrated in the region of
low momentum of the residual nuclear system $n{}^{10}$C, where the
quasi-free mechanism of the ${}^{12}$C$\left( {\gamma ,\, \pi ^{ -} p}
\right)^{11}$C$\quad $ reaction dominates.
\
In the case of the ${}^{12}$C$\left(
{\gamma ,\, \pi ^{ -} p} \right)p{}^{10}$B$\quad $ reaction, the
amplitude, corresponding to the diagram in Fig. 1\textit{d}, includes an
additional component, in which the proton of the $\pi ^{ -} p$ pair is a
nucleon of the $\Delta$\textit{N} system. This component of the amplitude in
properties is similar to the amplitudes of the second diagram group, which
have weak angular dependence. The contribution to the yield of the reaction
diagram in Fig. 1\textit{d} is shown in Fig. 4 by the dash-dot-dot curve.

The total contribution to the reaction yield of the $\pi ^{ -} p$ pair
produced in the $^{12}$C$\left( {\gamma ,\, \pi ^{ -} p}
\right)^{11}$C, $^{12}$C$\left( {\gamma ,\, \pi ^{ -} p}
\right)n^{10}$C, and $^{12}$C$\left( {\gamma ,\, \pi ^{ -} p}
\right)p{}^{10}$B reactions is shown in Fig. 4 by the solid curve. For
the ${}^{12}$C$\left( {\gamma ,\, \pi ^{ -} p} \right)n{}^{10}$C
reaction the mechanism, corresponding to the diagram in Fig. 1\textit{c},
dominates at the opening angle $\Theta _{\pi p} \simeq 180^\circ $. As can
be seen, for the large scattering angles calculated, the reaction yield is
an order of magnitude less than that of the experimental data.

\section{Conclusion}

We have presented a model of the pion photoproduction on the nucleus with
the emission of two nucleons in the$A\left( {\gamma ,\, \pi NN} \right)B$
reaction. In this model we have moved beyond the standard shell-model
considering $\Delta$\textit{N} correlations in the nuclear wave functions, which
are caused by the virtual transitions $NN \to \Delta N \to NN$ in the ground
state of the nucleus. The main ingredients of the model are the two- and
three-particle density matrixes and the transition operators $\gamma \Delta
\to N\pi $ and $\gamma N \to N\pi $. The direct and exchange reaction
mechanisms, which follow from the structure of the density matrices, were
examined. This model is an extension of our recent model for the $A\left(
{\gamma ,\, \pi N} \right)B$ reaction \cite{2} to the pion photoproduction in
the $A\left( {\gamma ,\, \pi NN} \right)B$ reaction.

The processes of pion production in the ${}^{12}$C$\left( {\gamma ,\,
\pi ^{ +} p} \right)$ and ${}^{12}$C$\left( {\gamma ,\, \pi ^{ -} p}
\right)$ reactions were considered. The analysis of these reactions is made
in the kinematic region of the large momentum transfers to the residual
nuclear system. The motivation for this work is based on the conclusion
drawn in \cite{2} that it is impossible to explain experimental data of the
${}^{12}$C$\left( {\gamma ,\, \pi ^{ +} p} \right)$ \cite{8} and
${}^{12}$C$\left( {\gamma ,\, \pi ^{ -} p} \right)$ \cite{7} reactions
assuming that the residual nuclei are in a bound state in these reactions.

The experimental data of the ${}^{12}$C$\left( {\gamma ,\, \pi ^{ +} p}
\right)$ reaction \cite{8} were explained satisfactorily by the total
contribution of the ${}^{12}$C$\left( {\gamma ,\, \pi ^{ +} p}
\right)^{11}$Be, ${}^{12}$C$\left( {\gamma ,\, \pi ^{ +} p}
\right)n^{10}$Be, and ${}^{12}$C$\left( {\gamma ,\, \pi ^{ +} p}
\right)p^{10}$Li reactions, using the reaction mechanisms due to
$\Delta$\textit{N}-correlations. It is shown that at the large momentum transfer
the photoproduction of the pions on the nucleus with the emission of two
nucleons is dominant. It was found that the $\left( {\gamma ,\, \pi
^{ +} p} \right)$ reaction is not very sensitive to the parameters of the
wave function of the $\Delta$\textit{N} system in the studied kinematic region, but
it allows us to estimate the probability of the virtual $NN \to \Delta N$
transition. In our model, the probability of the isobar configurations in
the $^{12}$C ground state per nucleon is 0.0126, which is consistent with
the experimental data of \cite{8}.

The results of an estimation of the isobar configuration contribution in the
cross-section of the ${}^{12}$C$\left( {\gamma ,\, \pi ^{ -} p}
\right)$ reaction are interesting in another aspect. The considered
mechanisms of the pion-nucleon pair production can be background or
imitative in relation to the processes that are accompanied by the
excitation of the hypothetical highly excited nucleus states \cite{10}. According
to our analysis, the cross-section of the ${}^{12}$C$\left( {\gamma ,\,
\pi ^{ -} p} \right)$ reaction, measured in the experiment \cite{7} at large
opening angles of the pion and proton, cannot be explained by the
contribution of the isobar configurations in the $^{12}$C ground state. So,
the question, brought up at different times in \cite{12,13,7,9,10}, about the
existence of the resonance and bound isobar-nuclear states that decay with
the emission of the pion-nucleon pair remains open.

\end{document}